\documentclass{iau} 
\usepackage{graphicx}

\title[Delays in WD evolution] %% give here short title %%
{Two delays in white dwarf evolution \\ revealed by {\it Gaia}}

\author[S. Cheng]   %% give here short author list %%
{Sihao Cheng$^1$
}
\affiliation{$^1$Department of Physics and Astronomy, The Johns Hopkins University, \\3400 N Charles Street, Baltimore, MD 21218, USA \\email: {\tt s.cheng@jhu.edu}} 
\pubyear{2019}
\volume{357}  %% insert here IAU Symposium No.
\jname{White Dwarfs as probes of fundamental physics and tracers of planetary, stellar \& galactic evolution}
\editors{A.C. Editor, B.D. Editor \& C.E. Editor, eds.}
\begin{document}

\maketitle

\begin{abstract}

By comparing two age indicators of high-mass white dwarfs derived from {\it Gaia} data, two discoveries have been made recently: one is the existence of a cooling anomaly that produces the Q branch structure on the Hertzsprung--Russell diagram, the other is the existence of double-white-dwarf merger products. The former poses a challenge for white dwarf cooling models, and the latter has implications on binary evolution and type-Ia supernovae.

\end{abstract}

\firstsection % if your document starts with a section,

\section{Introduction}

The unprecedented astrometric and photometric power of {\it Gaia} provides a unique opportunity to investigate white dwarf evolution.
On the one hand, the photometric isochrone age of a large sample of white dwarfs can be accurately derived from the Hertzsprung-Russell (H--R) diagram.
On the other hand, their kinematic information from {\it Gaia} makes it possible to infer the true age of these white dwarfs, according to the age--velocity-dispersion relation of the Milky Way disc. 
Because the photometric isochrone age is derived with a single-star evolution model and a standard white dwarf cooling model, any discrepancy between these two age indicators, or equivalently, the discrepancy between the observed velocity distribution and the velocity distribution predicted from the photoemtric ages and the age--velocity-dispersion relation, indicates deviations from the single-star evolution or the standard cooling model. Also, by carefully designing the comparison, one can distinguish more than one source of the deviation based on their different properties. In this article, I present two evolutionary delays of white dwarfs revealed by this method, which include an unexpected cooling delay and a merger delay from binary evolution, and discuss related astrophysical questions.

\section{Do we understand the cooling of white dwarfs?}

\begin{figure}
    \centering
    \includegraphics[width=0.7\columnwidth]{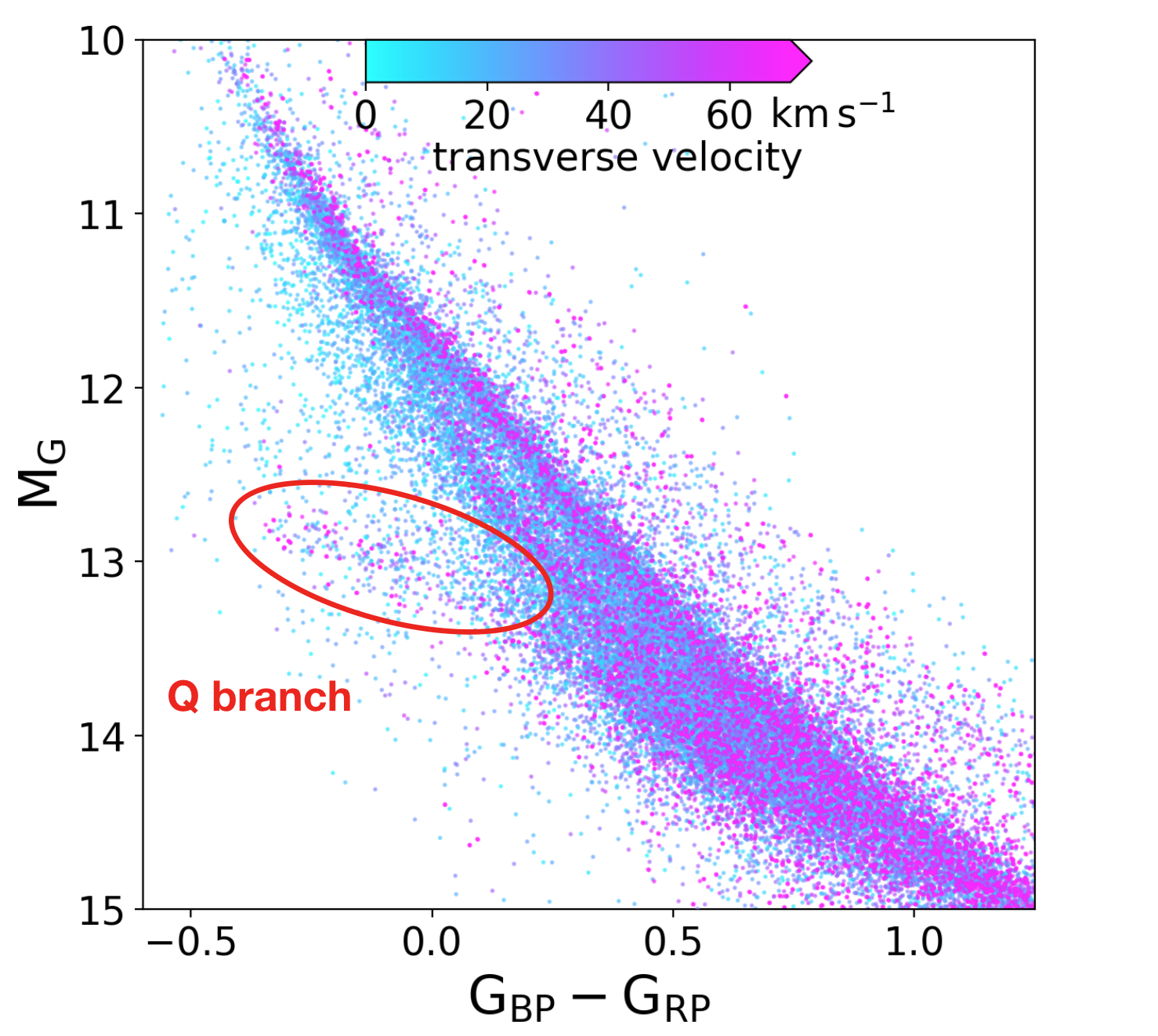}
    \caption{The H--R diagram of white dwarfs within 150 pc, colour-coded by transverse velocity. The data come from {\it Gaia} Data Release 2. Fast-moving high-mass white dwarfs are concentrated on the Q branch, which can only be explained by a significant slowing-down of cooling rate in a small fraction of white dwarfs.}
    \label{fig:WDHR}
\end{figure}

Most white dwarfs shine at the cost of losing their thermal energy and cooling down over cosmic time. Since the seminal work of \cite[Mestel (1952)]{Mestel1952}, a series of detailed cooling models have been developed, making white dwarfs accurate cosmic clocks (see e.g. \cite[Fontaine et al. 2001]{Fontaine2001} for a review).

Unexpectedly, the {\it Gaia} data (\cite[Gaia Collaboration et al. 2018]{GaiaCollaboration2018a}) revealed an over-density of high-mass white dwarfs on the H--R diagram, called the Q branch (Figure~\ref{fig:WDHR}). Standard white dwarf models do not anticipate this feature. An explanation was proposed by \cite[Tremblay et al. (2019)]{Tremblay2019} that the energy release from crystallisation produces this pile-up on the H--R diagram. However, \cite[Cheng et al. (2019a)]{Cheng2019a} found that on the Q branch, an anomalously high fraction of white dwarfs have a high transverse velocity (Figure 1). According to the age--velocity-dispersion relation of the Milky Way disc (e.g., \cite[Holmberg et al. 2009]{Holmberg2009}), these white dwarfs must be old. As argued by \cite[Cheng et al. (2019a)]{Cheng2019a}, different cooling behaviours are required to explain both the pile-up and the velocity distribution of high-mass white dwarfs (1.08--1.23 $M_\odot$): while most of them cool normally as the standard cooling model (which already includes crystallisation effect) predicts, about 6\% almost stop their cooling on the Q branch for as long as 8 billion years. Neither crystallisation delay or merger delay alone cannot explain all observations. So, we called this delay an `extra cooling delay' and the population the `extra delayed population'. This anomaly is both a challenge and an opportunity for understanding white dwarf cooling. 

To stop white dwarf cooling, an extra energy source is needed to provide the energy loss from shining. 
With a semi-analytical calculation, \cite[Cheng et al. (2019a)]{Cheng2019a} showed that the sink of $^{22}$Ne in C/O-core white dwarfs is a promising candidate for the extra energy source. In this explanation, the extra delayed white dwarfs are bizarre `sedimentars', as predicted and dubbed by \cite[Bildsten \& Hall (2001)]{Bildsten2001}; these white dwarfs shine out of the gravitational sedimentation of material instead of cooling. 

There are several aspects of this cooling anomaly and the Q branch worthy of further exploration. 

{\underline{\it More evidence for the extra cooling delay}}: Collecting more evidence for this cooling anomaly is still necessary. LP 93-21, a high-mass white dwarf just below the Q branch but with halo kinematics (\cite[Kawka et al. 2020]{Kawka2020}) may count as additional evidence. Other evidence may come from using asteroseismology to directly measure the cooling rate (as proposed by Bart Dunlap at this symposium) in the future.

{\underline{\it Cooling models with $^{22}$Ne settling}}: The $^{22}$Ne settling effect depends on the white dwarf mass, $^{22}$Ne abundance, core composition, crystallisation temperature, and the treatment of the physics of settling. Despite existing cooling models in the literature, there has not been a model incorporating both the $^{22}$Ne settling effect and the updated phase diagram of crystallisation for high-mass ($>1.1\,M_\odot$) C/O-core white dwarfs. Such models would test the $^{22}$Ne settling hypothesis and may provide new insight into white dwarf physics.

{\underline{\it DQ white dwarfs on the Q branch}}: All DQs on the Q branch belong to the extra delayed population, whereas among the extra delayed population, about half are DQs and the other half are DAs. Both the number density and carbon abundance suggest that the Q-branch DQs are evolutionary descendants of the hot-DQs discovered by \cite[Dufour et al. (2007)]{Dufour2007}. Further questions include the origin of the distinction between DQs and DAs and whether the Q-branch DQs also have magnetic field as the hot-DQs.

{\underline{\it Merger origin}}: Three arguments support a double-white-dwarf (double-WD) merger origin of the extra delayed white dwarfs: the $^{22}$Ne settling hypothesis for the delay (because it requires C/O-core), the possible relation to hot-DQs, and the lack of wide binary companions. Nevertheless, further definitive evidence is still needed.

{\underline{\it Core composition}}: We expect that the extra delayed white dwarfs have C/O cores instead of O/Ne cores that is expected for normal high-mass white dwarfs. Future observations that can determine the core composition will be crucial to testing both the $^{22}$Ne settling hypothesis and their merger origin.

\section{What are the origins of high-mass white dwarfs?}

High-mass white dwarfs can form from single-star evolution or from merger of stars. In particular, a considerable fraction of high-mass white dwarfs are expected to be double-WD merger products (e.g., \cite[Temmink et al. 2019]{Temmink2019}). However, despite discussions in the literature, there is still a lack of clear evidence for the white dwarfs that are merger products. The difficulty is to reliably distinguish the merger population from singly evolved white dwarfs.

The astrometric information from {\it Gaia} has improved the situation significantly. Because double-WD merger products experience additional binary evolution, their true ages are younger than their photometric isochrone ages. To get rid of the influence of the cooling anomaly described in the previous section, \cite[Cheng et al. (2019b)]{Cheng2019b} selected only high-mass white dwarfs above the Q branch and used their transverse velocity distribution to infer the fraction of merger products. Thanks to the much larger sample size than what \cite[Wegg \& Phinney (2012)]{Wegg2012} used in a previous study with similar method, \cite[Cheng et al. (2019b)]{Cheng2019b} found clear velocity excess that corresponds to about $20\pm6$\% of high-mass white dwarfs (0.8--1.3 $M_\odot$) being double-WD merger products. This is strong evidence for the merger origin of high-mass white dwarfs.

\section{How frequently do two white dwarfs merge?}

The fraction of double-WD merger products can also be translated into a double-WD merger rate in some mass range. Then, it can be compared with binary population synthesis results and other observations to improve our understanding of binary evolution, and it can also be compared with the type-Ia supernovae rate as double-WD merger is one of the promising scenarios of type-Ia supernovae. 

\begin{figure*}
    \centering
    \includegraphics[width=0.9\textwidth]{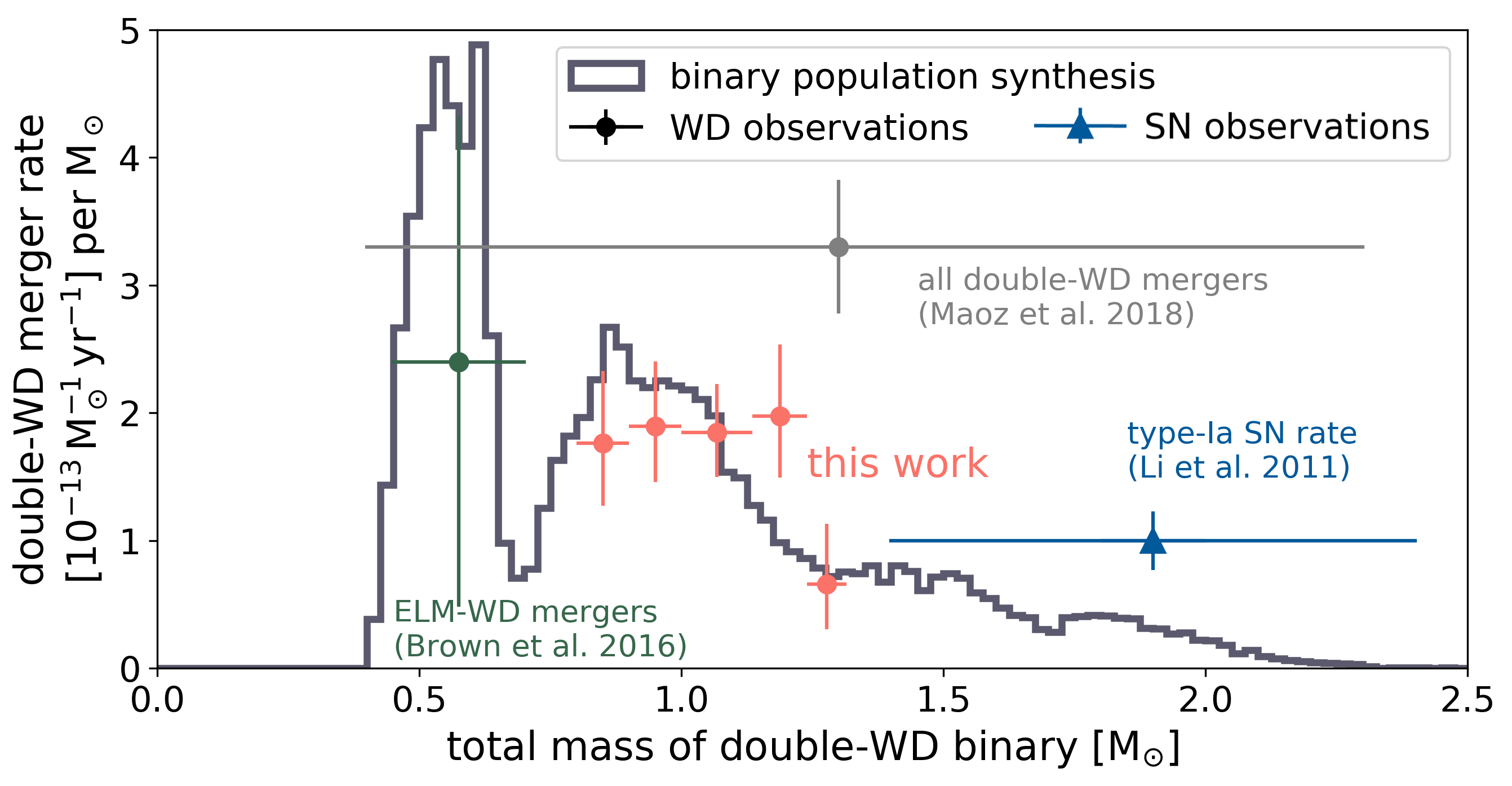}
    \caption{Estimates of the double-WD merger rate as a function of total mass, from both observational and theoretical sides. This figure is reproduced from the Figure 4 in \cite[Cheng et al. (2019b)]{Cheng2019b}.}
    \label{fig:mergerrate}
\end{figure*}

Here, I only discuss the double-WD mergers occurring in a close binary system. In general, there are three ways to measure the double-WD merger rate: to extrapolate from pre-merger binary systems, to count merger events, and to search for merger products. Previous works were mainly focused on extrapolating the orbital distribution of pre-merger systems (e.g., \cite[Brown et al. 2016, Maoz et al. 2018]{Brown2016, Maoz2018}), which currently provides no or low mass resolution and large uncertainties. The merger rate translated from the fraction of merger products in \cite[Cheng et al. (2019b)]{Cheng2019b} adds significant mass resolution and precision to previous measurements, and it is shown to be close to binary simulation results (Figure~\ref{fig:mergerrate}).

If it is allowed to extrapolate the observed merger rate to a higher mass range using the simulated mass distribution, one will conclude that the rate of mergers with total mass $>1.4\,M_\odot$ is a third to a half of the type-Ia supernova rate measured for Milky-Way-like galaxies (\cite[Li et al. 2011]{Li2011}). The fact that the two rates are close to each other supports the idea that massive double-WD mergers may contribute to a significant fraction of type-Ia supernovae, whereas the fact that the merger rate is lower than the supernova rate indicates other channels of type-Ia supernovae, including double-WD mergers with lower total mass and/or single degenerate progenitors.

\begin{thebibliography}{}

% \bibitem[Althaus et al. (2010)]{Althaus2010}{Althaus, L. G., Garc\'ia-Berro, E., Renedo, I., et al.} 2010, \textit{ApJ}, 719, 612

\bibitem[]{Bildsten2001}{Bildsten, L., \& Hall, D. M.} 2001, \textit{Ap. Lett.}, 549, L219

\bibitem[]{Brown2016}{Brown, W. R., Kilic, M., Kenyon, S. J., \& Gianninas, A.} 2016,
\textit{ApJ}, 824, 46

% \bibitem[]{Camisassa2016}{Camisassa, M. E., Althaus, L. G., C\'orsico, A. H., et al.} 2016, \textit{ApJ}, 823, 158

\bibitem[]{Cheng2019a}{Cheng, S., Cummings, J., \& M\'enard, B.} 2019a, \textit{ApJ}, 886, 100

\bibitem[]{Cheng2019b}{Cheng, S., Cummings, J., M\'enard, B., \& Toonen, S.} 2019b, arXiv:1910.09558

% \bibitem[]{Coutu2019}{Coutu, S., Dufour, P., Bergeron, P., et al.} 2019, \textit{ApJ}, 885, 74

\bibitem[]{Dufour2007}{Dufour, P., Liebert, J., Fontaine, G., \& Behara, N.} 2007, \textit{Nature}, 450, 522

\bibitem[]{Fontain2001}{Fontaine, G., Brassard, P., \& Bergeron, P.} 2001, \textit{PASP}, 113, 409

\bibitem[]{GaiaCollaboration2018}{Gaia Collaboration, Babusiaux, C., van Leeuwen, F., et al.} 2018, \textit{A\&A}, 616, A10

% \bibitem[]{GaiaCollaboration2016}{Gaia Collaboration, Prusti, T., de Bruijne, J. H. J., et al.} 2016, \textit{A\&A}, 595, A1

% \bibitem[]{Garcia-Berro2008}{Garc\'ia-Berro, E., Althaus, L. G., C\'orsico, A. H., \& Isern, J.} 2008, \textit{ApJ}, 677, 473

% \bibitem[]{Giammichele2012}{Giammichele, N., Bergeron, P., \& Dufour, P.} 2012, \textit{ApJS}, 199, 29

\bibitem[]{Holmberg2009}{Holmberg, J., Nordstr\"om, B., \& Andersen, J.} 2009, \textit{A\&A}, 501, 941

\bibitem[]{Kawka}{Kawka, A., Vennes, S., \& Ferrario, L.} 2019, \textit{MNRAS} Letters, 491, L40

\bibitem[]{Li2011}{Li, W., Chornock, R., Leaman, J., et al.} 2011, \textit{MNRAS}, 412, 1473

\bibitem[]{Maoz2018}{Maoz, D., Hallakoun, N., \& Badenes, C.} 2018, \textit{MNRAS}, 476, 2584

\bibitem[]{Mestel1952}{Mestel, L.} 1952, \textit{MNRAS}, 112, 583

\bibitem[]{Temmink2019}{Temmink, K. D., Toonen, S., Zapartas, E., Justham, S., \& G\"ansicke, B. T.} 2019, arXiv:1910.05335

% \bibitem[]{Toonen2017}{Toonen, S., Hollands, M., G\"ansicke, B. T., \& Boekholt, T.} 2017, \textit{A\&A},
% 602, A16

\bibitem[]{Tremblay2019}{Tremblay, P.-E., Fontaine, G., Fusillo, N. P. G., et al.} 2019, \textit{Nature}, 565, 202

\bibitem[]{Wegg2012}{Wegg, C., \& Phinney, E. S.} 2012, \textit{MNRAS}, 426, 427


\end{thebibliography}
\end{document}